\documentclass{emulateapj}
\usepackage{pstricks,apjfonts}

\def\chandra{{\em Chandra}\/}
\def\asca{{\em ASCA}\/}
\def\vla{{\em VLA}\/}
\def\rosat{{\em ROSAT}\/}
\def\xmm{{\em XMM}\/}
\def\1e{\mbox{1E\,0657--56}}

\def\gax{\gtrsim}

\def\arcsec{$^{\prime\prime}$}
\def\arcmin{$^{\prime}$}

\begin{document}

\submitted{ApJ, in press}

\lefthead{TEMPERATURE MAPS FOR RADIO HALO CLUSTERS} 

\righthead{GOVONI ET AL.}

\title{{\em CHANDRA}\/ TEMPERATURE MAPS FOR GALAXY CLUSTERS WITH RADIO
HALOS}

\author{F.~Govoni$^{1,2}$, M.~Markevitch$^3$, A.~Vikhlinin$^3$,
L.~VanSpeybroeck $^{3,4}$, L.~Feretti$^1$, and G.~Giovannini$^{1,2}$}

\affil{Full-resolution version at http://hea-www.harvard.edu/$\sim$maxim/papers/rhalos\_tmaps} 

\altaffiltext{1}{Istituto di Radioastronomia del CNR, Via Gobetti 101, 
  I-40129 Bologna, Italy}
\altaffiltext{2}{Dipartimento di Astronomia, Universit\`a di Bologna, 
  Via Ranzani 1, I-40127 Bologna, Italy}
\altaffiltext{3}{Harvard-Smithsonian Center for Astrophysics, 60 Garden St.,
  Cambridge, MA 02138}
\altaffiltext{4}{We honor the memory of our late colleague}

\begin{abstract}

We analyze \chandra\ temperature maps for a sample of clusters with
high-quality radio halo data, to study the origin of the radio halos.  The
sample includes A520, A665, A754, A773, A1914, A2163, A2218, A2319, and \1e.
We present new temperature maps for all but two of them (A520 and A754).
All these clusters exhibit distorted X-ray morphology and strong gas
temperature variations indicating ongoing mergers.  Some clusters, e.g.,
A520, A665, \1e, exhibit the previously reported spatial correlation between
the radio halo brightness and the hot gas regions.  However, it is not a
general feature.  While most mergers are too messy to allow us to
disentangle the projection effects, we find clear counterexamples (e.g.,
A754 and A773) where the hottest gas regions do not exhibit radio emission
at the present sensitivity level.  This cannot be explained by projection
effects, and therefore argues against merger shocks --- at least those
relatively weak ones responsible for the observed temperature structure in
most clusters --- as the main mechanism for the halo generation.  This
leaves merger-generated turbulence as a more likely mechanism.  The two
clusters with the clearest radio brightness -- temperature correlation, A520
and \1e, are both mergers in which a small dense subcluster has just passed
through the main cluster, very likely generating turbulence in its wake.
The maximum radio brightness and the hot gas are both seen in these wake
regions.  On the other hand, the halos in \1e\ and A665 (both high-velocity
mergers) extend into the shock regions in front of the subclusters, where no
strong turbulence is expected.  Thus, in high-velocity ($M\simeq 2-3$)
mergers, both shock and turbulence acceleration mechanisms may be
significant.

\end{abstract}

\keywords{ galaxies: clusters: general --- galaxies: clusters: individual
(A520, A665, A754, A773, A1914, A2163, A2218, A2319, \1e) --- intergalactic
medium --- X-rays: galaxies: clusters --- Radio continuum}

\section{INTRODUCTION}

The baryonic content of the galaxy clusters is dominated by the hot ($T\sim
2-10$ keV) intergalactic gas whose thermal emission is observed in X-rays.
Many clusters also exhibit diffuse radio sources that have no apparent
connection to any individual cluster galaxy.  In this paper, we are
interested in the extremely low-brightness, large-scale sources classified
as {\it radio halos}. 
They often span the whole cluster and are unpolarized.
Other diffuse radio sources, not considered here, include strongly elongated
{\em relics}, 
which are highly polarized and often found at the cluster
periphery, and {\it mini-halos}\/ which are found around the powerful
central radio galaxy; for recent reviews see e.g. Kempner et al. (2003),
Feretti (2003).

While the cluster X-ray emission is due to thermal electrons with energies
of several keV, the radio halo emission at $\sim 1$ GHz is produced by
synchrotron radiation of relativistic electrons with energies of $\sim 10$
GeV in magnetic fields with $B\simeq 0.5-1\;\mu$G. These electrons should
coexist with the thermal population.  Their origin is still uncertain; the
difficulty in explaining their presence arises from the combination of the
large sizes of halos ($r\sim 1$ Mpc) and the short synchrotron lifetime of
these electrons ($10^{7}-10^{8}$ yrs).  One needs a mechanism by which these
electrons are locally and simultaneously (re-)accelerated over the halo
volume. Several such mechanisms of feeding energy to the relativistic
electrons have been proposed (see, e.g., recent reviews by En{\ss}lin 1999;
Sarazin 2001; Brunetti 2002; Petrosian 2002 and references therein). We will
briefly review theoretical arguments in \S\ref{sec:disc}.

On the observational side, the number of known radio halos is already
sufficient to start looking for correlations with other cluster properties
in order to elucidate their acceleration mechanism.  Halos are typically
found in clusters with significant substructure in the X-ray brightness
which indicates merger activity (e.g., Feretti 1999; Buote 2001).  Halos are
present in rich clusters, characterized by high X-ray luminosities and
temperatures.  The percentage of clusters with halos in a complete X-ray
flux-limited sample (that includes systems with $L_X> 5\times 10^{44}$ erg
s$^{-1}$ in the $0.1-2.4$ keV band) is $\simeq 5$\%.  The halo fraction
increases with the X-ray luminosity, to $\simeq 33$\% for clusters with
$L_X> 10^{45}$ erg s$^{-1}$ (Giovannini, Feretti, \& Govoni 1999a).  The radio power of
a halo, if one is present, strongly correlates with the cluster luminosity,
gas temperature (e.g., Liang et al.\ 2000; Colafrancesco 1999; Feretti
1999), or total mass (Govoni et al.\ 2001a).  In a number of well-resolved
clusters, a spatial correlation between the radio halo and X-ray brightness
is observed (Govoni et al.\ 2001b).  These observations indicate that radio
halos are closely related to the intra-cluster thermal gas, its history and
energetics. However, details of this connection need further investigations.

Because the halos appear to be related to cluster mergers, cluster gas
temperature maps, which contain information on the merger geometry, stage,
and velocity, can provide further information on their origin.  Indeed,
\chandra\ studies of A2163, A665 (Markevitch \& Vikhlinin 2001, hereafter
MV01), \1e\ (Markevitch et al. 2002a, hereafter M02a) 
and A520 (Markevitch et al.\ 2002b, hereafter M02b) revealed a possible
spatial correlation between the high-temperature cluster regions and the
diffuse radio emission.  A665 showed a spatial correlation between the radio
halo brightness and the location of a possible bow shock, appearing to
support a merger shock origin for the relativistic halo electrons.

Clearly, an X-ray and radio study of a greater cluster sample is needed to
see whether a spatial correlation between the radio halo and the gas
temperature is a common feature.  We present \chandra\ X-ray images and
temperature maps for such a sample.

We make the radio -- X-ray comparison for the clusters A520, A665, A754,
A773, A1914, A2163, A2218, A2319, and \1e.  \chandra\ maps for A754 and A520
are reproduced from Markevitch et al.\ (2003a) and M02b, respectively.  For
A2163 and \1e, we present temperature maps from new observations, deeper
than those analyzed by MV01 and M02a; the full derivation of these maps will
be presented by Markevitch et al.\ (in prep.)  A higher-accuracy temperature
map for A665 is derived in this work by adding a new long observation to the
one used in MV01.  For A773, A1914, A2218, and A2319, we derive the first
high-resolution temperature maps here.

The radio data were taken from previously published work (A2319: Feretti,
Giovannini, \& B\"ohringer 1997; A665, A2218: Giovannini \& Feretti 2000;
\1e: Liang et al.\ 2000; A520, A773: Govoni et al.\ 2001a; A2163: Feretti et
al.\ 2001; A754, A1914: Bacchi et al.\ 2003).

We use $H_0=70$ km s$^{-1}$ Mpc$^{-1}$, $\Omega_0=0.3$ and
$\Omega_\Lambda=0.7$.

\begin{table}
\begin{center}
\renewcommand{\tabcolsep}{2mm}
\renewcommand{\arraystretch}{1.3}
\caption{X-Ray data}
\begin{tabular}{p{2cm}crrl}
\hline
\hline
Name       & $z$       & ObsID    & exposure,  & Observation \\
           &           &          &  ksec &     date       \\ 
\hline
 A520\dotfill      & 0.1990    & 528      & 9.3    & 2000 Oct 10  \\  
 A665\dotfill      & 0.1819    & 531      & 8.5    & 1999 Dec 29  \\
                   &           & 3586     & 23.8   & 2002 Dec 28 \\
 A754\dotfill      & 0.0542    & 577      & 39.1   & 1999 Oct 30  \\     
 A773\dotfill      & 0.2170    & 533      & 10.4   & 2000 Sep  5  \\
                   &           & 3588     &  8.3   & 2003 Jan 25  \\
 A1914\dotfill     & 0.1712    & 3593     & 17.6   & 2003 Sep 3 \\
 A2163\dotfill     & 0.2030    & 1653     & 69.6   & 2001 Jun 16  \\
 A2218\dotfill     & 0.1756    &  553     &  5.2   & 1999 Oct 19    \\
                   &           & 1454     & 10.9   & 1999 Oct 19    \\
                   &           & 1666     & 22.8   & 2001 Aug 30    \\
 A2319\dotfill     & 0.0557    & 3231     & 14.0   & 2002 Mar 15    \\
\1e\dotfill        & 0.296$^a$ & 3184     & 69.6   & 2002 Jul 12 \\ 
\hline
\multicolumn{5}{l}{\scriptsize Col. 1: Cluster name;}\\
\multicolumn{5}{l}{\scriptsize Col. 2: Redshift (Struble \& Rood 1999;
 $^a$ Tucker et al.\ 1998)}\\
\multicolumn{5}{l}{\scriptsize Col. 3: \chandra\ observation ID number;}\\ 
\multicolumn{5}{l}{\scriptsize Col. 4: Clean exposure time;}\\
\multicolumn{5}{l}{\scriptsize Col. 5: Observing date.}\\ 
\label{chandra}
\end{tabular}
\vspace{-5mm}
\end{center}
\end{table}

\section{X-RAY ANALYSIS}

We used archival and our proprietary \chandra\ ACIS data to derive X-ray
images and gas temperature maps. ACIS%
\footnote{http://asc.harvard.edu/cal}
has an energy band of 0.3--9 keV and combines $1''$ angular resolution and a
moderate energy resolution.  We used the {\sc CIAO}%
\footnote{http://cxc.harvard.edu/ciao}
software package for data processing, {\sc XSPEC} and A. Vikhlinin's {\sc
ZHTOOLS} to fit the overall spectra and analyze images, and our own code to
derive the temperature maps.

The X-ray observations are identified by their ID numbers (ObsID) in Table
\ref{chandra}, where the clean exposure time and the observing date are
listed.  All observations were performed with the ACIS-I detector which
covers a field of view of 16\arcmin$\times$16\arcmin.

Except for A1914, A2163, and \1e, for which we use the longer of the two
available exposures (of very different length), when more than one
observation existed for the same object, the observations were combined.

\subsection{X-ray data preparation}

For each observation, standard screening was applied to the photon list.  We
removed bad pixels and columns, events with {\em ASCA}\/ grades 1,5 and 7,
and cosmic ray afterglows.
The background was modeled using a composite blank-field background dataset
corresponding to the period of the observations, cleaned and normalized as
described by Markevitch et al.\ (2003b).  In particular, we first excluded
flare periods, extracting the X-ray light curves from the source-free parts
of the field and requiring that the 0.3--12 keV background rate be within a
factor of 1.2 of the nominal background.  After the flare exclusion, we
normalized the blank-sky background dataset to the cluster observation by
the ratio of counts at high energies (in the PHA interval of 2500--3000 ADU;
approximately 10--12 keV) which is free from the sky emission. This
correction was always small, within 10\% of the exposure ratio.  If an
observation was performed in VF mode, and if the corresponding VF-mode
background dataset existed (ObsIDs 1666, 1653, 3184, 3231, 3586, 3588,
3593), additional particle background reduction (Vikhlinin 2001) was applied
to the cluster and background data.

The soft Galactic background in the direction of the analyzed clusters
(Snowden et al.\ 1997) is not much different than in the regions included in
the blank-sky background --- so no special treatment was required --- for
all clusters but A2163 and A2319.  For A2163, treatment of the soft Galactic
excess is described in MV01; the central region of A2319 that we analyze
here is so bright that such an excess does not matter.  In general, our
temperature maps are limited to the brightest regions of the clusters where
the background accuracy is not critical.

\subsection{X-ray spectra}

For consistency checks, we first derived the average temperatures of our
clusters, extracting spectra from circular regions of $r=0.8$ Mpc (or
smaller if such a region did not fit within the field of view). This radius
encompasses most of the cluster emission. Point sources were excluded.  The
corresponding spectral Redistribution Matrix Files (RMF) and Auxiliary
Responce Files (ARF) were generated using the A. Vikhlinin's software and
were weighted by the cluster brightness. The large quantum efficiency (QE)
spatial non-uniformity caused by CTI in the data prior to Feb 2000 (with the
ACIS focal plane $-110$C) was modeled using the formula from Vikhlinin
(2000); for other data, the standard QE uniformity maps were used.  ARFs for
the ACIS-I observations also included the time- and position-independent
fudge factor of 0.93 at $E<1.8$ keV to account for the backside / frontside
illuminated CCD flux discrepancy in the current calibration.  The
time-dependent, position-independent correction for the soft absorption
caused by contamination buildup (Plucinsky et al.\ 2003) was included in the
ARFs.

Spectra were fit in the $0.8-10$ keV energy band with the MEKAL model
(Kaastra 1992) by fixing the absorption column to the Galactic value (Dickey
\& Lockman 1990). Two exceptions are A2163 and \1e\ where it is known to be
significantly different, higher for A2163 (e.g., Elbaz, 
Arnaud, \& B\"ohringer, 1995) and
lower for \1e\ (e.g., Liang et al.\ 2000). For A2163, it was fit as a free
parameter, while for \1e, the Liang et al.'s best-fit \rosat\ value was
used, which is in good agreement with our data.

The resulting average temperatures and abundances, at the 90\% confidence,
are given in Table 2 together with the extraction radii and absorption
columns.  All the obtained temperature are in good agreement with previously
derived values found in the literature.

\subsection{Temperature maps}

Temperature images of the cluster central regions were derived as described
in Markevitch et al.\ (2000a) and MV01. Specifically, we excluded point
sources, extracted images in several energy bands, subtracted the
background, divided the resulting images by the exposure maps, and smoothed
them by a Gaussian filter with variable width (same in all bands) to get
useful statistical accuracy over the interesting brightness range.  A
temperature in each pixel of the map was obtained by fitting values in each
pixel of these images with a thermal plasma, fixing $N_H$ to the Galactic
(or other, see above) value and the element abundance to 0.3 solar.  To
verify the resulting temperature maps, we fit spectra from several
interesting regions of each cluster using {\sc XSPEC}. The outlying areas of
the temperature maps, where noise starts to dominate (usually where the
$1\sigma$ errors are greater than $\sim 2$ keV), were cut off.

\begin{table}
\begin{center}
\renewcommand{\tabcolsep}{2mm}
\renewcommand{\arraystretch}{1.3}
\caption{Fits to average cluster spectra}
\begin{tabular}{p{2cm}rccc}
\hline
\hline
Name   & $T$,           & abund.        &  radius,   & $N_H,$               \\
       & keV            &               &  \arcsec & $10^{20}$ cm$^{-2}$ \\ 
\hline
A520\dotfill   & 7.1 $\pm$0.7     & 0.24$\pm$0.12 & 245       & 7.80       \\
A665\dotfill   & 8.2 $\pm$0.5     & 0.27$\pm$0.07 & 260       & 4.24       \\
A754\dotfill   & 10.0$\pm$0.3     & 0.30$\pm$0.05 & 540       & 4.36      \\
A773\dotfill   & 7.5 $\pm$0.8     & 0.34$\pm$0.11 & 230       & 1.44      \\
A1914\dotfill  & 10.9$\pm$0.7     & 0.24$\pm$0.08 & 275       & 0.95      \\
A2163\dotfill  & 12.4$\pm$0.7     & 0.20$\pm$0.03 & 240       & $18.7\pm3.5$\\
A2218\dotfill  &  6.7$\pm$0.5     & 0.10$\pm$0.07 & 270       &3.24    \\
A2319\dotfill  & 10.1$\pm$0.4     & 0.30$\pm$0.05 & 540       &7.93    \\
\1e\dotfill    & $13.9\pm0.7$     & 0.25$\pm$0.05 & 180       &4.60     \\
\hline
\multicolumn{5}{l}{\scriptsize Col. 1: Cluster name;}\\ 
\multicolumn{5}{l}{\scriptsize Col. 2: Cluster temperature;}\\
\multicolumn{5}{l}{\scriptsize Col. 3: Abundance relative to solar;}\\ 
\multicolumn{5}{l}{\scriptsize Col. 4: Radius of the spectrum extraction
  region}\\
\multicolumn{5}{l}{\scriptsize Col. 5: Galactic absorption column (Dickey \& Lockman 1990),}\\
\multicolumn{5}{l}{\scriptsize except for A2163 and \1e, see text.}\\ 
\label{tempe}
\end{tabular}
\vspace{-5mm}
\end{center}
\end{table}

\section{Results}

Below we present our X-ray images and temperature maps, discuss how they
elucidate the merger state and geometry, and compare them to the radio halo
images.  For each cluster, we present a temperature map with an X-ray and a
radio contour plot overlaid, an optical image with an X-ray contour plot
overlaid, and an X-ray image with a radio contour plot overlaid.

\subsection{Abell 520}

\begin{figure*}[t]

\includegraphics[scale=1.0]{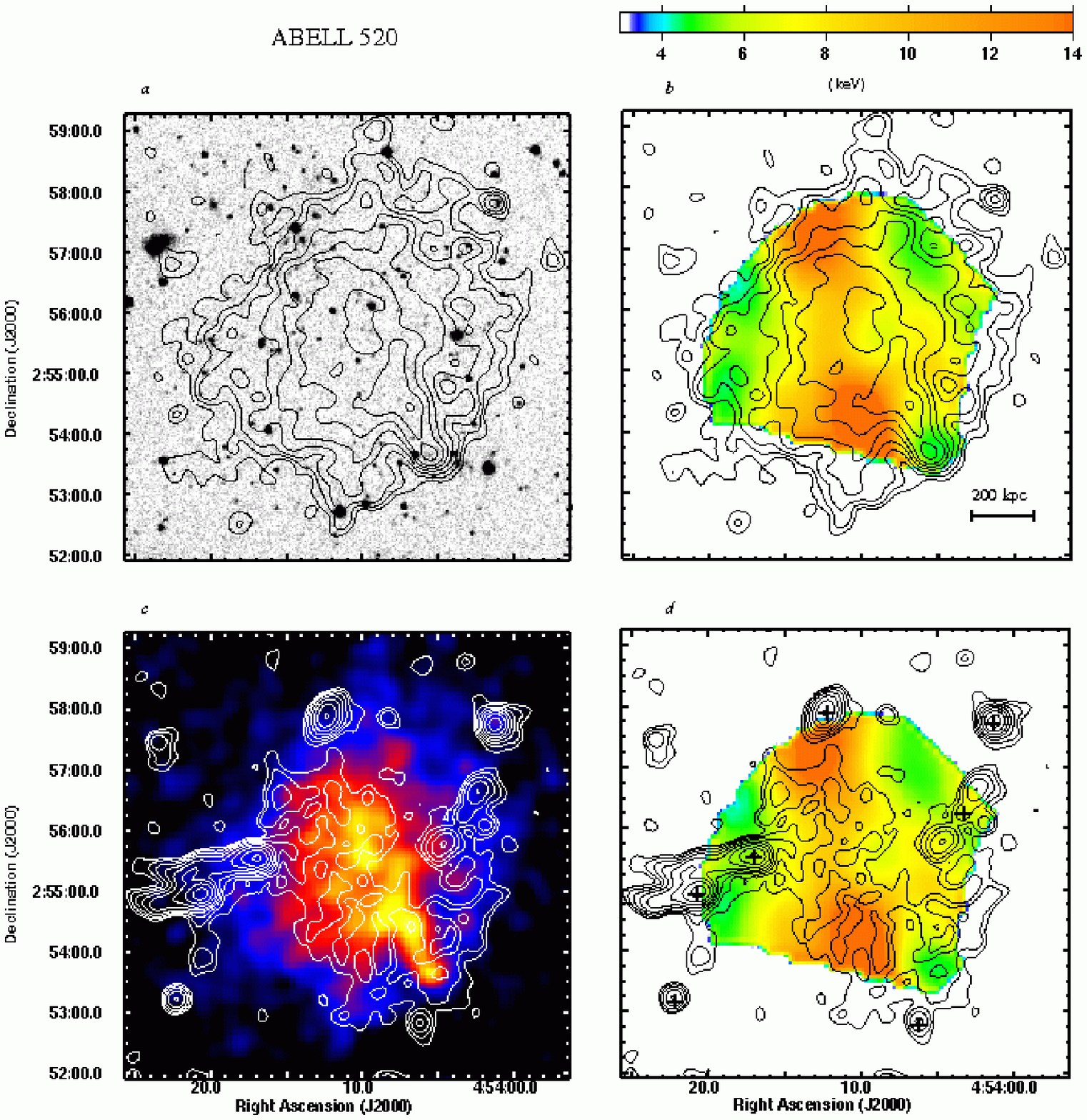}

\caption{A520. 
({\em a}\/):  X-ray contours overlaid on the optical DSS image.  
  The $0.8-4$ keV X-ray image is adaptively smoothed; contours are
  spaced by a factor of $\sqrt2$. 
({\em b}\/): X-ray contour plot overlaid on the temperature map (colors).  
({\em  c}\/): The isocontour map at 1.4 GHz
  of the central region of A520 overlaid on the X-ray image (colors). The
  radio image has a FWHM of $15''\times15''$. The contour levels are: 
  0.06, 0.12, 0.24, 0.48, 0.96, 1.92, 3.84, 7.68 mJy/beam.
({\em d}\/): Radio contours overlaid on the temperature map (colors); 
  crosses mark some radio sources unrelated to the halo emission. 
}
\label{fig:A520}
\end{figure*}

Optical spectroscopic data on A520 suggest that this cluster is undergoing
strong dynamical evolution (Proust et al.\ 2000).  A short \chandra\
observation (Table 1) placed this X-ray-luminous cluster in the ACIS chip
I3.  The X-ray observation confirms that A520 is in the middle of a merger.
In Fig.\ \ref{fig:A520}, we compare an X-ray image and a temperature map
(from M02b) derived from this \chandra\ observation, a \vla\ radio halo
image at 1.4 GHz from Govoni et al.\ (2001a),
and an optical Digitized Sky Survey (DSS)%
\footnote{http://archive.stsci.edu/dss}
plate.

The most prominent feature in the X-ray data is a dense, compact --- but
clearly extended --- cool gas clump southwest of the center.  Apparently, it
has just passed straight through the main cluster from the northeast,
accompanying a possible group of galaxies (although not centered on any one
galaxy).  The temperature map shows that the bright gas trail it has left
behind is cool, and reveals a hot strip of apparently shock-heated gas along
this trail.  Most of the radio halo emission appears to follow this hot
strip, and not the cool bright trail.  The X-ray image hints at a possible
bow shock in front of the dense clump, coincident with the southwest edge of
the radio halo, but a longer observation (planned for late 2003) is needed
to determine the nature of that feature.  This cluster is one of the best
examples of a spatial correlation between the radio halo brightness and the
gas temperature.  It is also one of only two clusters in our sample with a
simple and clear merger geometry (the other being \1e).

\subsection{Abell 665}

The X-ray image of A665 strongly suggests that the cluster is undergoing a
merger (e.g., Jones \& Saunders 1996; G{\'o}mez, Hughes, \& Birkinshaw 2000).  The merger
apparently occurs in the direction of the elongation of the galaxy
distribution (Geller \& Beers 1982; Beers \& Tonry 1986).  Indeed, a
\chandra\ temperature map revealed large variations and a possible shock
with $M\approx 2$ in the expected location (MV01).  A luminous radio halo in
A665 was discovered by Moffet \& Birkinshaw (1989), confirmed by Jones \&
Saunders (1996) and further studied by Giovannini \& Feretti (2000).

\begin{figure*}[t]

\includegraphics[scale=1.0]{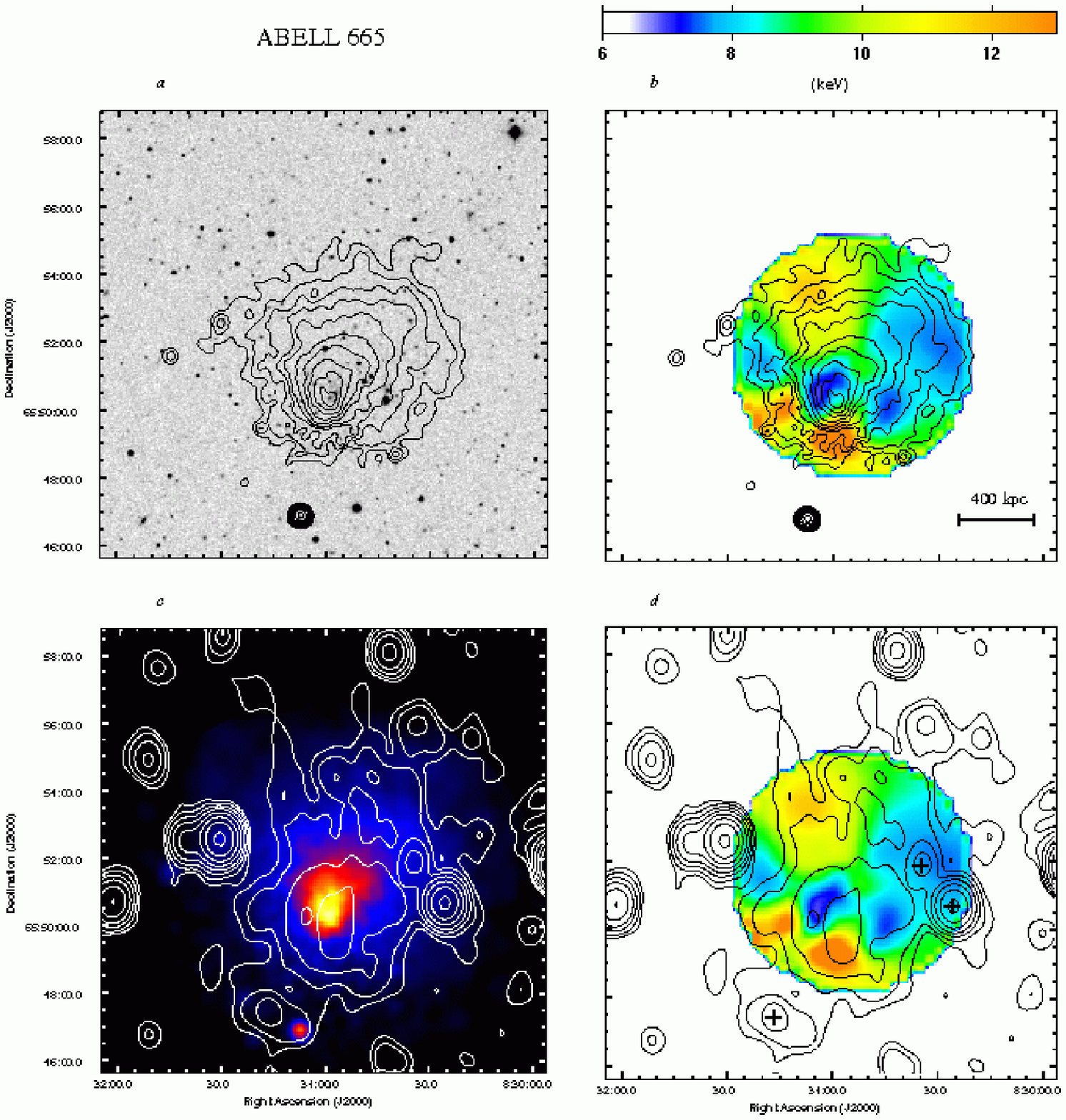}

\caption{A665.
({\em a}\/):  X-ray contours overlaid on the optical DSS image.  
  The $0.8-4$ keV X-ray image is adaptively smoothed; contours are
  spaced by a factor of $\sqrt2$. 
({\em b}\/): X-ray contour plot overlaid on the temperature map (colors). 
({\em c}\/): The isocontour map at 1.4 GHz
  of the central region of A665 overlaid on the X-ray image (colors). The
  radio image has a FWHM of $52''\times42''$. The contour levels are: 
  0.2 0.4 0.8 1.5 3 6 12 25 mJy/beam.
({\em d}\/): Radio contours overlaid on the temperature map (colors); 
  crosses mark some radio sources unrelated to the halo emission. 
}
\label{fig:A665}
\end{figure*}

The cluster was observed by \chandra\ twice (see Table 1).  Here we derive a
more accurate temperature map than that presented in MV01, by including the
recent longer exposure.  In Fig.\ \ref{fig:A665}, the resulting map and an
ACIS image are compared to the 1.4 GHz \vla\ image (Giovannini \& Feretti
2000) and an optical DSS plate.  The new map confirms the presence of hot
--- most probably shock-heated --- gas south or in front of the cool
core. The new image, however, still does not show any corresponding density
feature as sharp as the bow shock observed in \1e. The edge of the cool core
is also not as sharp as in some other clusters, so the merger probably
proceeds at an angle to the sky plane.

The X-ray brightness peak is offset from the central galaxy, indicating gas
motion in the core.  Indeed, the new temperature map reveals complex
temperature structure inside the core as well as in its wake, including two
streams (or trails) of cool gas in the NW and NE directions from the core,
and a hot stream north of the core.

The radio halo is very extended and elongated in the SE-NW direction, which
is the apparent merger direction and the X-ray elongation.  As noted by
MV01, the ``leading edge'' of the halo extends beyond the cool core and
coincides with the shock region.  MV01 presented an image of the ratio of
the radio brightness to the square root of the X-ray brightness, which has
the approximate meaning of the density of relativistic electrons to thermal
electrons (assuming a uniform magnetic field).  Its maximum coincided with
the southern shock region.

\subsection{Abell 754}

\begin{figure*}[t]

\includegraphics[scale=1.0]{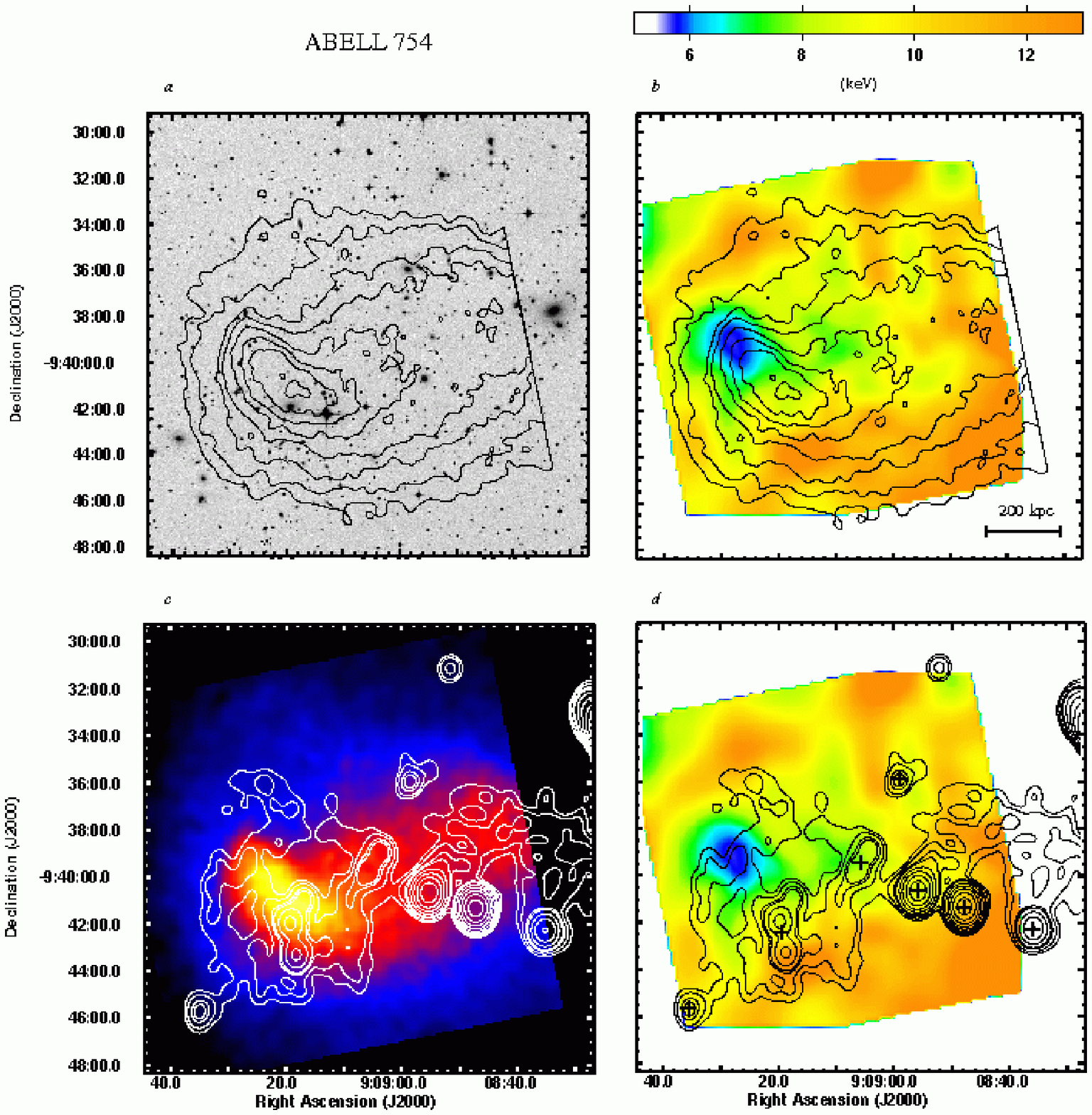}

\caption{A754. 
({\em a}\/):  X-ray contours overlaid on the optical DSS image.  
  The $0.8-5.5$ keV X-ray image is adaptively smoothed; contours are
  spaced by a factor of $\sqrt2$. 
({\em b}\/): X-ray contour plot overlaid on the temperature map (colors). 
({\em c}\/): The isocontour map at 1.4 GHz
  of the central region of A754 overlaid on the X-ray image (colors). The
  radio image has a FWHM of $55''\times44''$. The contour levels are: 
  0.45 0.8 1.5 2 4 8 16 32 64 mJy/beam.
({\em d}\/): Radio contours overlaid on the temperature map (colors); 
  crosses mark some radio sources unrelated to the halo emission. 
}
\label{fig:A754}
\end{figure*}

Optical and X-ray studies (e.g., Fabricant et al.\ 1986; Bird 1994; Slezak,
Durret, \& Gerbal 1994; Zabludoff \& Zaritsky 1995; Henry \& Briel 1995;
Henriksen \& Markevitch 1996; Bliton et al.\ 1998; De Grandi \& Molendi
2001) showed that A754 is undergoing a violent merger.  Markevitch et al.\
(2003a) noted that it is difficult to explain the complex details revealed
by the \chandra\ X-ray image and temperature map (which we reproduce here in
Fig.\ \ref{fig:A754}) with a simple two-cluster merger model.  Krivonos et
al.\ (2003) noted a possible weak bow shock east of the bright core in the
\rosat\ image; it is located approximately at the leftmost X-ray brightness
contour in Fig.\ \ref{fig:A754}.

A radio halo in A754 was discovered by Harris, Kapahi, \& Ekers (1980)
and recently
confirmed by Kassim et al.\ (2001) through radio observation at 0.3 GHz.  A
hard X-ray excess at $E\gax 45$ keV with respect to thermal emission was
also reported (Fusco-Femiano et al.\ 2003). 
The most likely explanation is inverse Compton emission from
the same relativistic electrons responsible for the diffuse radio emission,

In Fig.\ \ref{fig:A754}, we compare the \chandra\ X-ray image and
temperature map with a DSS optical plate and a 1.4 GHz \vla\ image from
Bacchi et al.\ (2003).  The radio diffuse emission is complex and very
extended.  Bacchi et al.\ (2003) noted that it consists of two large
components, roughly coincident with the two main galaxy concentrations.  The
temperature map shows strong spatial variations, such as a large hot area in
the south and southwest and cool gas at the tip of the bright tongue-like
structure.  This structure appears to be the remains of the core of one of
the subclusters, presently flowing in the northeast direction.  The eastern
diffuse radio emission is located around the eastern interface between the
tongue and the surrounding gas. The western radio emission extends beyond
the boundary of our temperature map.  That region was covered by the \asca\
map (Henriksen \& Markevitch 1996) which showed hot gas; more details will
soon be provided by \xmm\ (P. Henry \& A. Finoguenov, private
communication).  Significantly, the large, hot southern region in the
present map does not exhibit radio emission at the present sensitivity
level.

\subsection{Abell 773}

\begin{figure*}[t]

\includegraphics[scale=1.0]{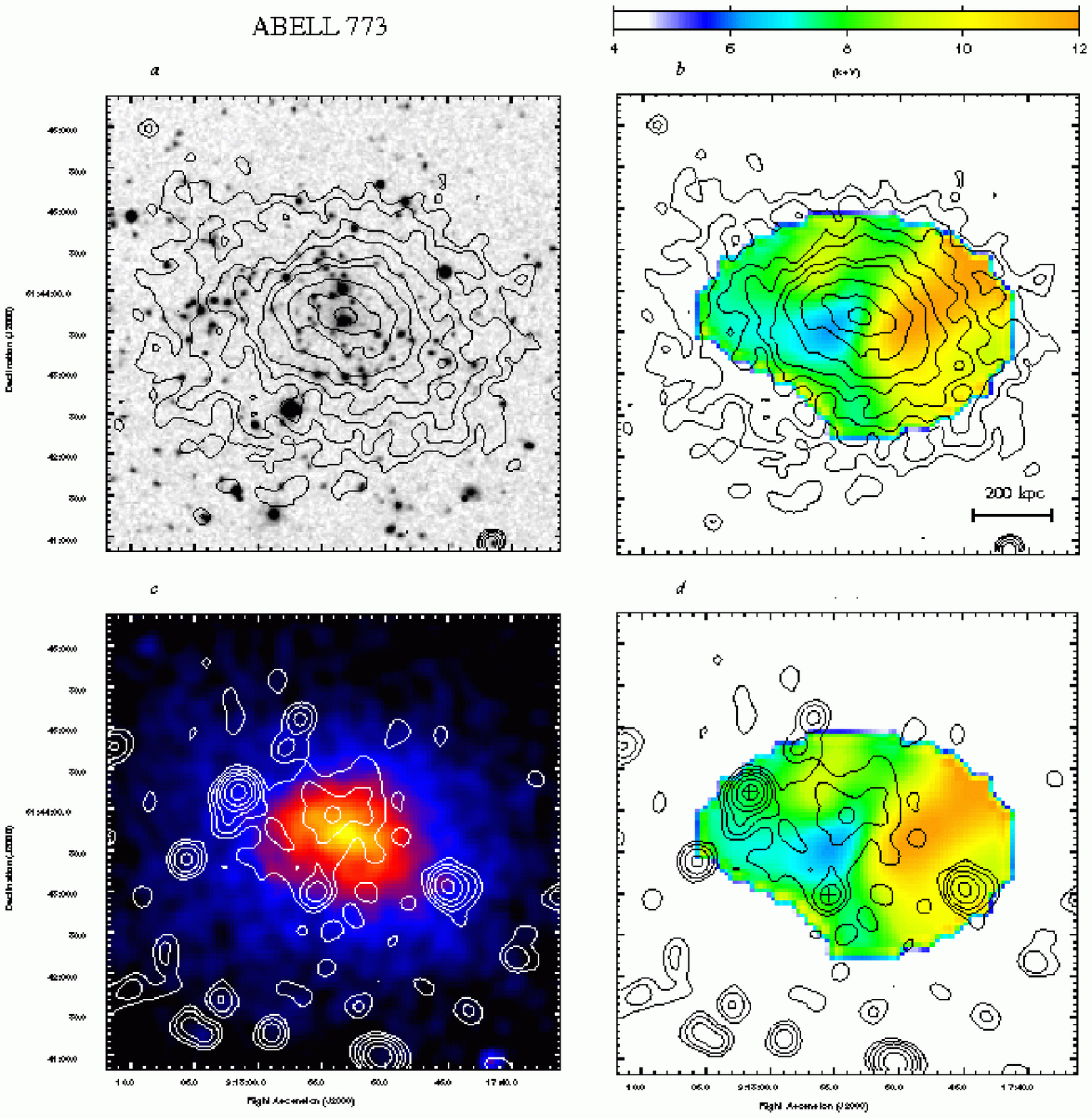}

\caption{A773. 
({\em a}\/):  X-ray contours overlaid on the optical DSS image.  
  The $0.8-4$ keV X-ray image is adaptively smoothed; contours are
  spaced by a factor of $\sqrt2$. 
({\em b}\/): X-ray contour plot overlaid on the temperature map (colors). 
({\em c}\/): The isocontour map at 1.4 GHz
  of the central region of A773 overlaid on the X-ray image (colors). The
  radio image has a FWHM of $15''\times15''$. The contour levels are: 
0.06 0.12 0.24 0.48 0.96 1.92 3.84 mJy/beam.
({\em d}\/): Radio contours overlaid on the temperature map (colors); 
  crosses mark some radio sources unrelated to the halo emission. 
}
\label{fig:A773}
\end{figure*}

An irregular X-ray shape of A773 was noticed in the \rosat\ data (e.g.,
Pierre \& Starck 1998; Rizza et al.\ 1998; Govoni et al.\ 2001a).  A radio
halo in this cluster was suggested by Giovannini, Tordi, \& Feretti (1999b) 
from the
NRAO \vla\ Sky Survey (NVSS, Condon et al.\ 1998) and confirmed by Govoni et
al.\ (2001a) through a dedicated \vla\ observation at 1.4 GHz.  In Fig.\
\ref{fig:A773}, we compare this radio image with an optical DSS plate, and
our temperature map and \chandra\ image.  This is the first temperature map
for A773.  It reveals strong temperature variations in the $6-12$ keV range.

The optical image clearly shows two galaxy subclusters, one in the center of
the X-ray emission and another at the eastern outskirt. Taking into account
the temperature map, it appears as thought the eastern galaxy group is
currently exiting the merger site, having shed its gas due to ram pressure
at the entry into the main cluster at its southwest side. The collision of
the gas clouds has generated a shock-heated area there, seen in the
temperature map.

The radio halo does not follow the X-ray brightness, nor temperature
distribution -- moreover, the hottest cluster region is not at all bright in
the radio (although a more sensitive, lower spatial resolution radio image
shows a faint extension mostly southward; Govoni et al.\ 2001a). Instead,
the radio halo is centered in the relatively cool region between the two
galaxy subclusters.

\subsection{Abell 1914}

The presence of a radio halo in A1914 was suggested by Giovannini et al.\
(1999b) from an NVSS search.  It was detected by Kempner \& Sarazin (2001) in
the Westerbork Northern Sky Survey (WENSS) at 0.3 GHz, and confirmed by
Bacchi et al.\ (2003) in deep \vla\ observations at 1.4 GHz.

\begin{figure*}[t]

\includegraphics[scale=1.0]{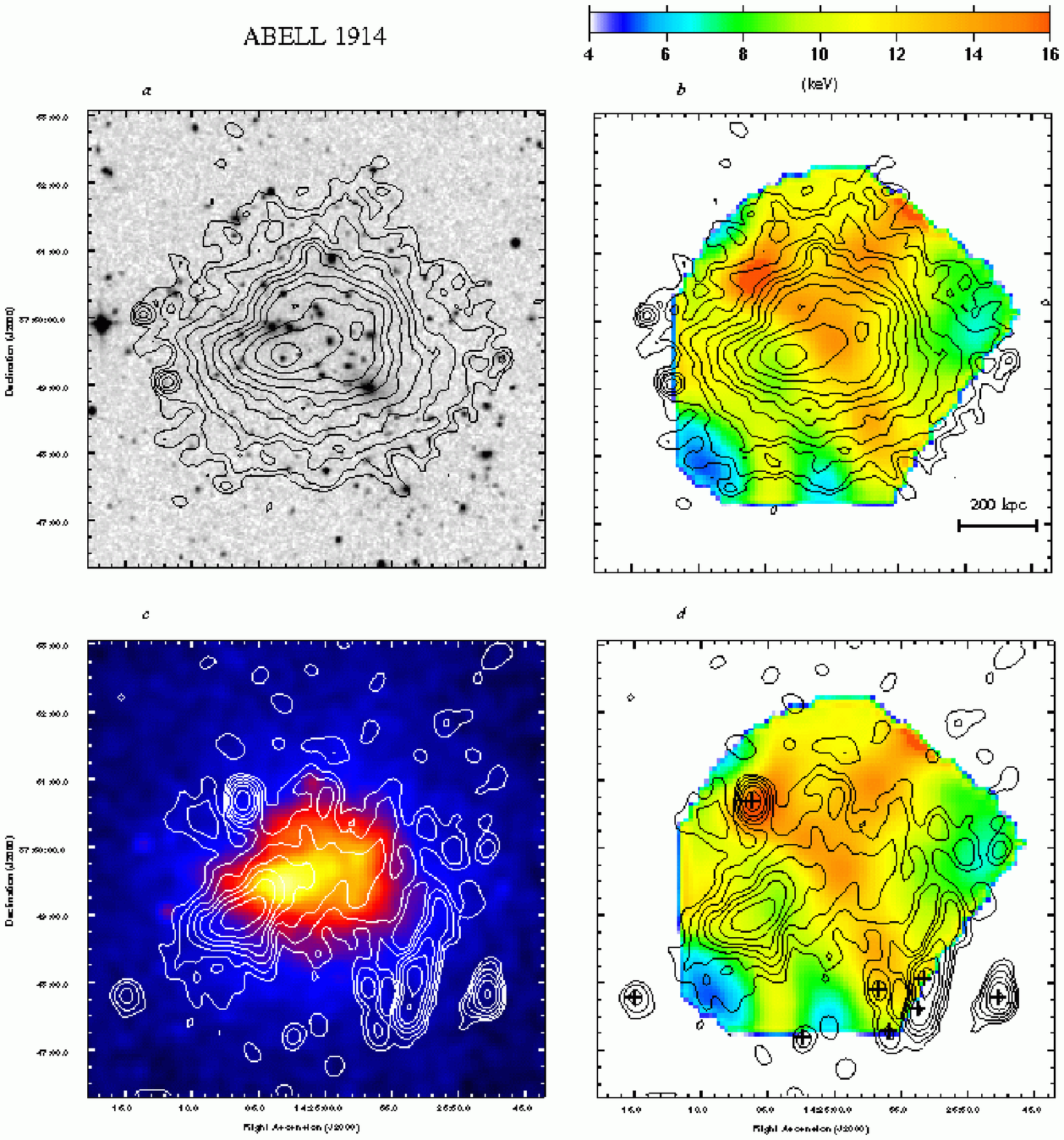}

\caption{A1914. 
({\em a}\/):  X-ray contours overlaid on the optical DSS image.  
  The $0.8-4$ keV X-ray image is adaptively smoothed; contours are
  spaced by a factor of $\sqrt2$. 
({\em b}\/): X-ray contour plot overlaid on the temperature map (colors). 
({\em c}\/): The isocontour map at 1.4 GHz
  of the central region of A1914 overlaid on the X-ray image (colors). The
  radio image has a FWHM of $20''\times15''$. The contour levels are: 
0.07 0.14 0.28 0.56 1 2 4 8 16 mJy/beam.
({\em d}\/): Radio contours overlaid on the temperature map (colors); 
  crosses mark some radio sources unrelated to the halo emission. 
}
\label{fig:A1914}
\end{figure*}

A1914 was observed by \chandra\ twice, but because of the relatively short
useful exposure of the first observation (ObsID 542), only the second one is
used here.  In Fig.\ \ref{fig:A1914}, we compare the radio image from Bacchi
et al.\ (2003) with our \chandra\ X-ray image and temperature map, and a DSS
plate.  This temperature map, the first for A1914, immediately provides a
likely merger scenario, not at all obvious from the X-ray and optical images
alone.  In this scenario, the NE-SW arch-like hot region through the cluster
center, which coincides with a distinct component in the X-ray image, is a
shock between the two infalling subclusters. One of them has arrived from
the southeast, where the map shows cool gas probably stripped from that
subcluster.  Its gas core was partly shocked and stopped by the collision
around the position of the present X-ray brightness peak. This gas is
currently squirting sideways along the hot arch, creating the prominent
cooler elongation to the east. Its galaxies appear to have penetrated
further, inside the shock-heated gas, and are now seen as a northeastern
galaxy concentration.  The brightest cluster galaxy was probably at the
center of the other colliding subcluster, which arrived from the west.  Its
cooler core was left behind and is currently seen as a western cool
extension, some of it possibly squirting along the western side of the
shocked region, similarly to the situation on the southeastern side of the
cluster.  If this merger had a smaller impact parameter and occurred closer
to the sky plane, we would now see a classical picture from the simulations
with two subclusters just before core passage and a shocked region between
them (e.g., Schindler \& M\"uller 1993 and many later works).

The diffuse radio emission in A1914 is unusual in that it has a distinct,
bright, elongated region, approximately along the presumed path traveled by
the eastern subcluster. It is accompanied by a more typical, diffuse,
low-brightness halo in the cluster center.  The bright feature is clearly
extended (Bacchi et al.\ 2003).  It would be interesting to determine if
this region is physically distinct from the rest of the halo.  An upper
limit on its polarization is 3\% (Bacchi et al.\ 2003), which excludes the
possibility that it is a relic projected onto the cluster center.  The
remaining low-brightness region of the radio halo approximately coincides
with the hot central region of the X-ray cluster and may even follow the
presumed streams of the gas of the two subclusters.  However, a detailed
comparison requires better accuracy and a removal of the radio sources in
the south and the resolution of the nature of the bright feature.

\subsection{Abell 2163}

A2163 is among the hottest and most X-ray luminous clusters known (e.g.,
Arnaud et al.\ 1992).  The presence of an extended, powerful radio halo in
A2163 was first reported by Herbig \& Birkinshaw (1994).  The \rosat\ image
(e.g., Elbaz et al.\ 1995), crude \asca\ temperature map (Markevitch et al.\
1994) and the previously published \chandra\ temperature map (MV01) clearly
show that the cluster is a merger.  However, even with the temperature map,
its geometry is difficult to determine.  

\begin{figure*}[t]

\includegraphics[scale=1.0]{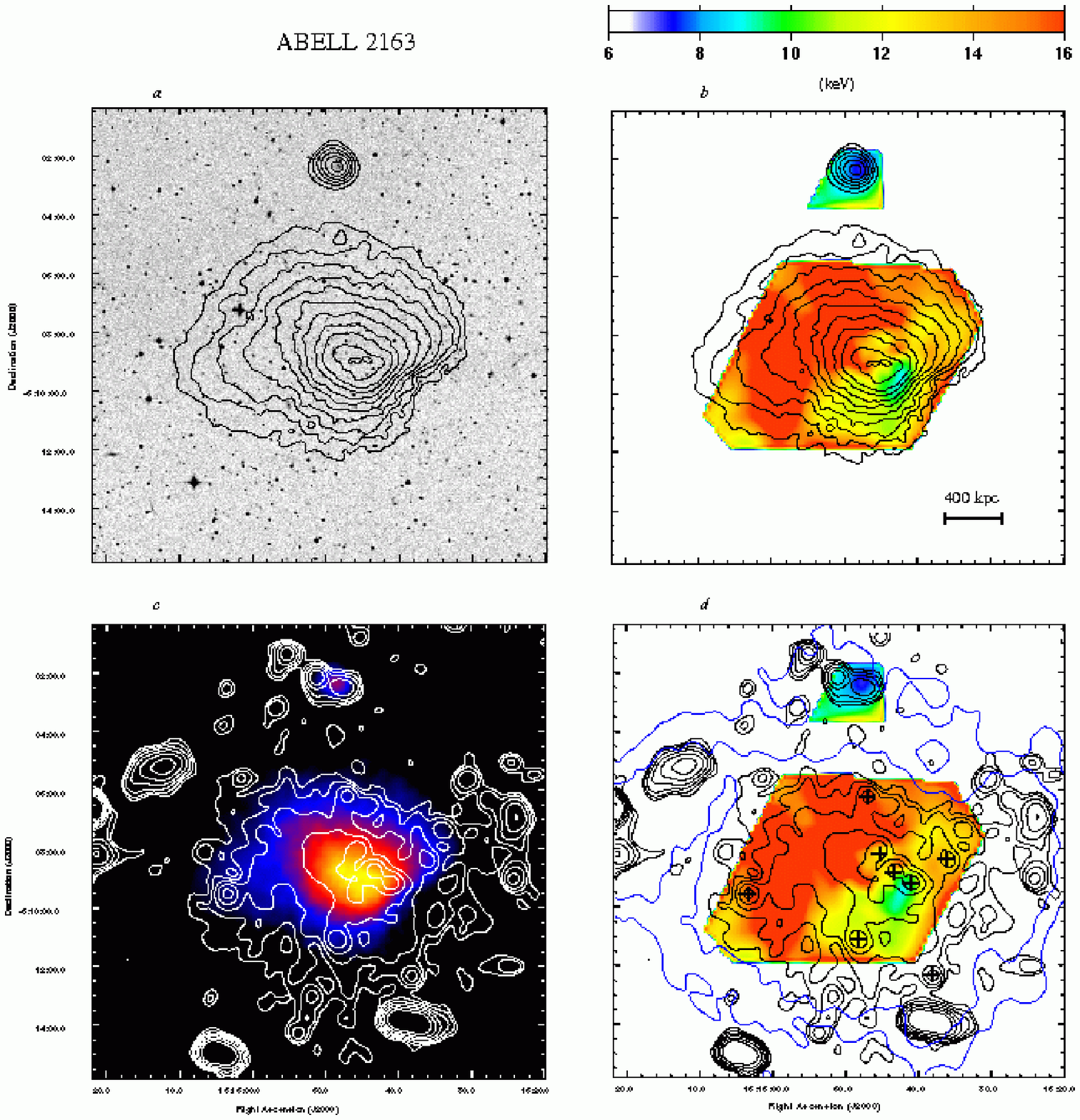}

\caption{A2163. 
({\em a}\/): X-ray contours overlaid on the optical DSS image.  The
$0.8-6.5$ keV X-ray image is adaptively smoothed; contours are spaced by a
factor of $\sqrt2$.  ({\em b}\/): X-ray contour plot overlaid on the
temperature map (colors).  ({\em c}\/): The isocontour map at 1.4 GHz of the
central region of A2163 overlaid on the X-ray image (colors). The radio
image has a FWHM of $30''\times30''$. The contour levels are: 0.1 0.2 0.4
0.8 1.6 3.2 mJy/beam.  ({\em d}\/): Radio contours overlaid on the
temperature map (colors); crosses mark some radio sources unrelated to the
halo emission. In this plate, we show two additional radio contours from a
lower-resolution (FWHM $45''\times60''$, blue outer contours).}
\label{fig:A2163}
\end{figure*}

In Fig.\ \ref{fig:A2163}, we present a new temperature map derived from a
deeper \chandra\ exposure by Markevitch et al.\ (in preparation). It is in
agreement with but more accurate than the MV01 map.  The X-ray image and the
new map show streams of hot and cold gas flowing in different directions, as
well as a possible remnant of a cool gas core.  It still does not clarify
the geometry of the merger, however.  This, the absence of any sharp
features in the high-statistics X-ray image, and the optical spectroscopic
data (Soucail et al., in preparation) all point to the possibility that the
merger is occurring at a large angle to the sky plane.

The temperature map is compared with an optical DSS plate and the radio halo
image from Feretti et al.\ (2001). The latter combines two images with
different FWHM resolutions, $30''$ (inner black contours) and the more
sensitive $45''\times60''$ (outer blue contours), respectively, to show the
large extent of the radio halo. The halo extension $2-3'$ east of center
coincides with the hottest region of the cluster; however, as noted in MV01,
this region also coincides with the X-ray brightness extension. In general,
the halo brightness follows the distribution of the X-ray brightness quite
well (Ferretti et al.\ 2001).

\subsection{Abell 2218}

\begin{figure*}[t]

\includegraphics[scale=1.0]{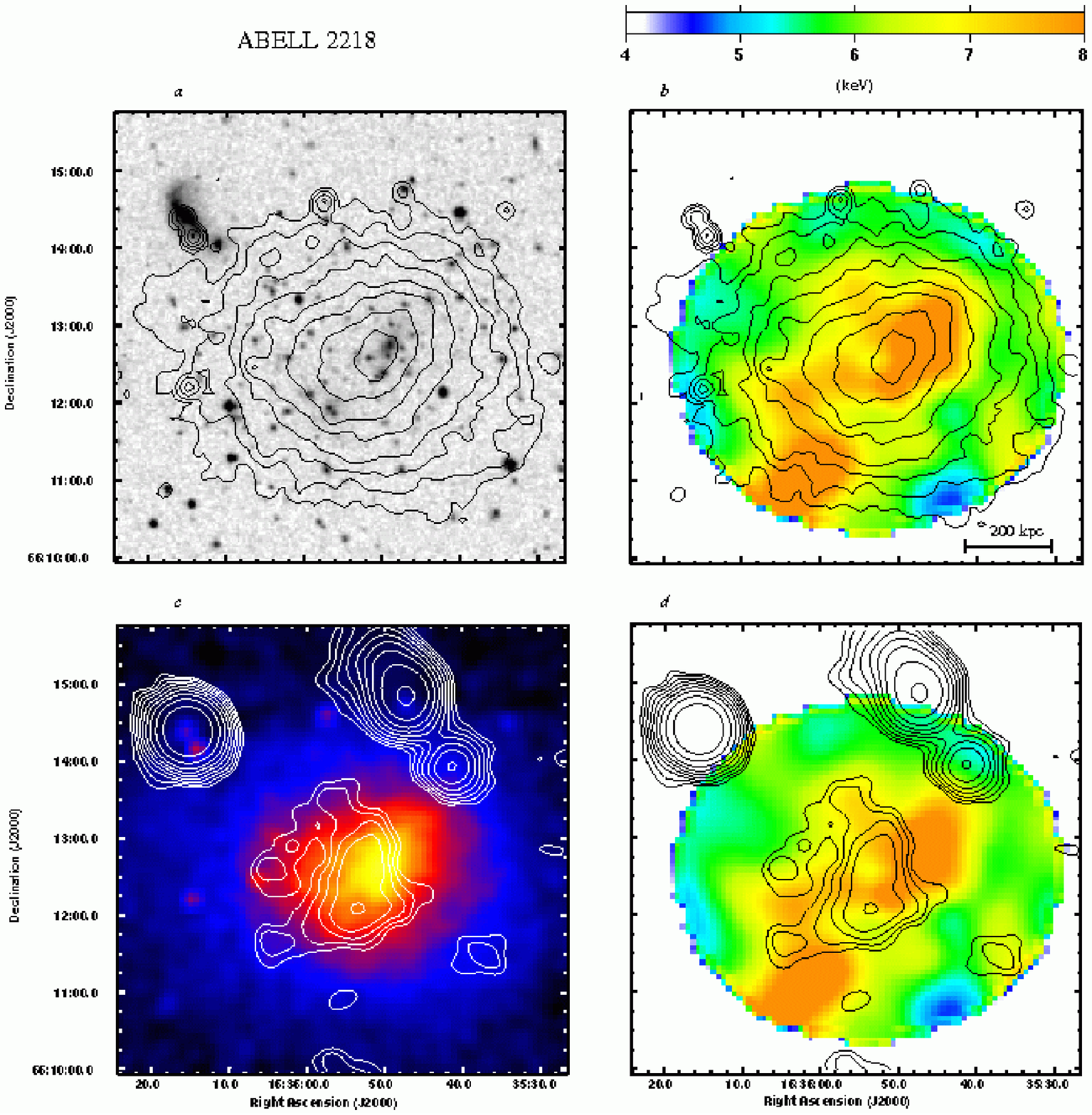}

\caption{A2218. 
({\em a}\/):  X-ray contours overlaid on the optical DSS image.  
  The $0.8-4.0$ keV X-ray image is adaptively smoothed; contours are
  spaced by a factor of $\sqrt2$. 
({\em b}\/): X-ray contour plot overlaid on the temperature map (colors). 
({\em c}\/): The isocontour map at 1.4 GHz
  of the central region of A2218 overlaid on the X-ray image (colors). The
  radio image has a FWHM of $35''\times35''$. The contour levels are: 
0.24 0.34 0.48 0.68 0.96 1.35 1.92 3 5 mJy/beam.
({\em d}\/): Radio contours overlaid on the temperature map (colors). 
}
\label{fig:A2218}
\end{figure*}

In A2218, strong and weak lensing mass reconstruction revealed two distinct
mass peaks inside the cluster core, approximately around the two brightest
galaxies along the NW-SE direction seen in the optical plate Fig.\
\ref{fig:A2218}{\em a}\/ (e.g., Abdelsalam, Saha, \& Williams 1998; Squires
et al.\ 1996; Kneib et al.\ 1995, 1996).  \rosat\ images suggested that the
cluster is not relaxed (e.g., Kneib et al.\ 1995; Squires et al.\ 1996;
Markevitch 1997), although at large scales, it appeared relatively
symmetric.

The cluster was observed with \chandra\ three times. Analysis of the first
two exposures was presented by Markevitch et al.\ (2000b) who found a central
peak in the radial temperature profile, and Machacek et al.\ (2002) who
reported an azimuthally asymmetric temperature structure in the core.  Here
we add a more recent, longer exposure and derive the first detailed
temperature map of A2218. It is shown in Fig.\ \ref{fig:A2218} along with
the X-ray image, a 1.4 GHz \vla\ image from Giovannini \& Feretti (2000), and
an optical DSS plate.

The map confirms the results from the earlier \chandra\ analyses and reveals
strong asymmetric temperature variations in the $5-10$ keV range.  
Together with the relatively symmetric X-ray image, such an irregular but
centrally peaked temperature map suggests that A2218 is at a later merger
stage, when the violent gas motions are starting to subside, as seen, e.g.,
in the simulations by Roettiger, Stone, \& Mushotzky (1998).

The relatively small radio halo in A2218 (Giovannini \& Feretti 2000) is
slightly (by about $40''$) offset from the X-ray brightness peak and does
not show a particularly strong correlation with the temperature map --- in
fact, the hottest spot in the cluster is not seen in the radio, as in A754
and A773.  The radio image has a relatively low sensitivity, however.

\subsection{Abell 2319}

\begin{figure*}[t]

\includegraphics[scale=1.0]{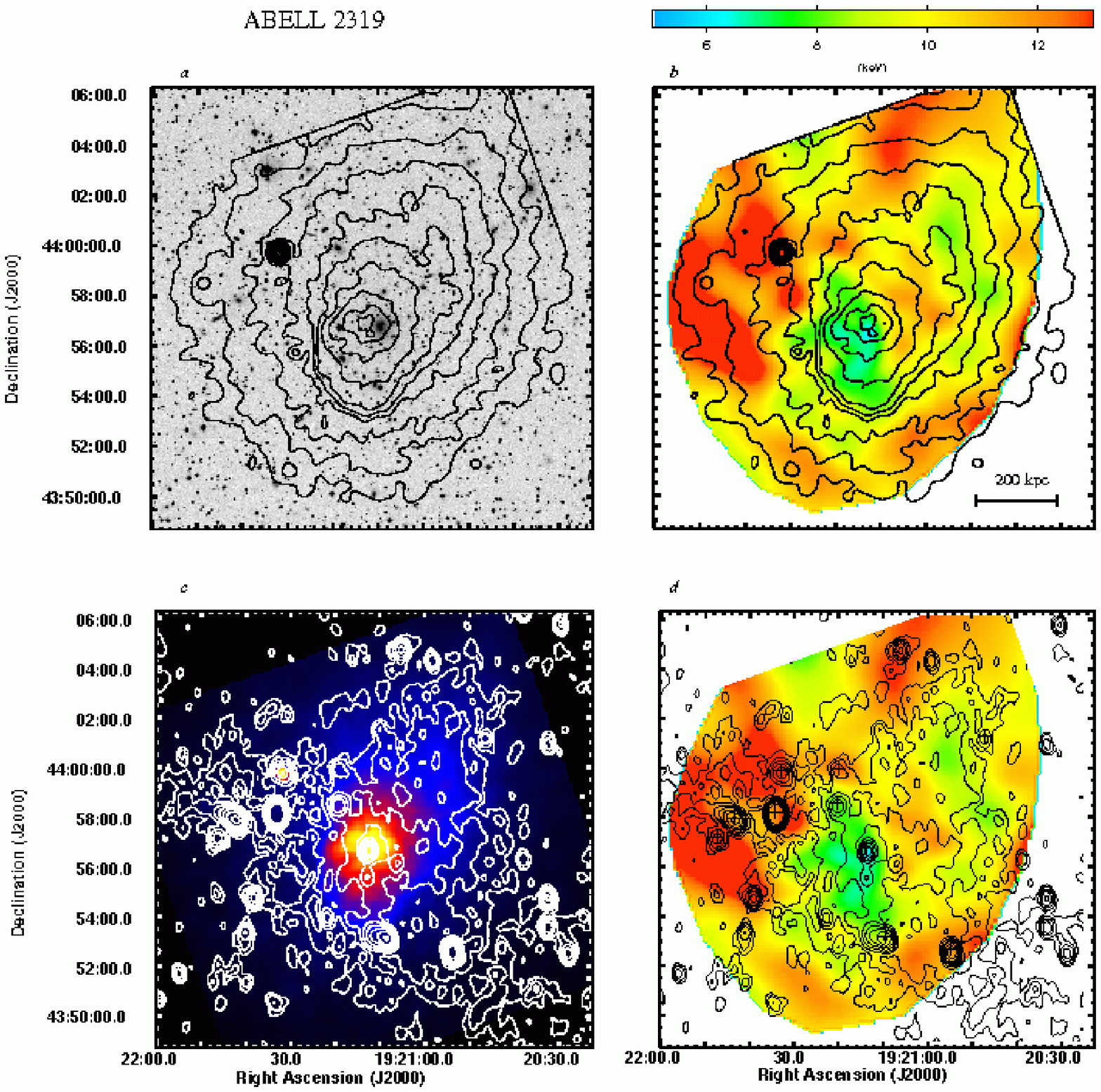}

\caption{A2319. 
({\em a}\/):  X-ray contours overlaid on the optical DSS image.  
  The $0.8-4.0$ keV X-ray image is adaptively smoothed; contours are
  spaced by a factor of $\sqrt2$. 
({\em b}\/): X-ray contour plot overlaid on the temperature map (colors). 
({\em c}\/): The isocontour map at 1.4 GHz
  of the central region of A2319 overlaid on the X-ray image (colors). The
  radio image has a FWHM of $29.0''\times20.4''$. The contour levels are: 
0.1 0.2 0.4 0.8 1.6 3.2 6.4 12.8 mJy/beam.
({\em d}\/): Radio contours overlaid on the temperature map (colors); 
  crosses mark some radio sources unrelated to the halo emission. 
}
\label{fig:A2319}
\end{figure*}

Optical analyses of the bright nearby cluster A2319 (e.g., Faber \& Dressler
1977, Oegerle, Hill, \& Fitchett 1995, Tr{\` e}vese, Cirimele, \& De Simone 2000) 
suggested that it consists
of two components superimposed along the line of sight, with a subcluster
around the second-brightest galaxy projected about $10'$ northwest of the cD
galaxy (seen in Fig.\ \ref{fig:A2319}{\em a}). That subcluster is seen as a
cool X-ray extension in the \rosat\ image and the crude \asca\ temperature
map (Markevitch 1996).  A2319 exhibits an extended and powerful radio halo
(Harris \& Miley 1978) with an irregular morphology well correlated with the
X-ray brightness (Feretti et al.\ 1997).

We use a \chandra\ observation of A2319 to derive the first detailed
temperature map of this cluster, shown in Fig.\ \ref{fig:A2319} along with
our X-ray image and a 1.4 GHz \vla\ image from Feretti et al.\ (1997).  The
most prominent feature seen in the X-ray data is a sharp cold front
southeast of the cD galaxy, such as those discovered by \chandra\ in many
other merging clusters (e.g., Markevitch et al.\ 2000a; Vikhlinin, Markevitch
\& Murray 2001). The central cool gas cloud is clearly moving southeast with
respect to the ambient hotter gas (or, equivalently, the hotter gas is
flowing around it to the northwest). The cD galaxy is neither at the
centroid nor at the coolest spot of the cluster, suggesting that the cool
gas core is moving independently of this galaxy. The cD galaxy appears to
have its own gas density peak on a smaller linear scale, itself displaced
eastward from the galaxy. (We note that this brightness peak falls exactly
on the ACIS-I central low-exposure spot, so the apparent temperature dip
there may be an artifact.)  We can also see a cool arm extending around the
cluster center from the tip of the cold front in the general direction of
the northwestern subcluster.  It may either be a tail of that subcluster, or
gas stripped from the cold front cloud, unrelated to the subcluster.
Overall, the picture suggests a later stage of a merger, well past the
initial encounter.

The radio halo follows remarkably closely the distribution of the cool gas
in the core, except for two low-brightness extensions into the hotter gas
northeast and southwest of the cold front.  The radio halo is more extended
southwest of the X-ray brightness centroid --- toward the cooler gas that we
observe there.

\subsection{\1e}

\1e\ is one of the hottest, most luminous clusters known (Tucker et al.\
1998) which also contains the most luminous radio halo (Liang et al.\ 2000).
An early \chandra\ observation revealed a spectacular bow shock propagating
in front of a dense, cool bullet-like subcluster exiting the site of the
collision with a bigger subcluster (M02a).  Those authors
derived an approximate temperature map of this merger and noticed that the
radio halo brightness peaks at the hottest cluster region.

\begin{figure*}[t]

\includegraphics[scale=1.0]{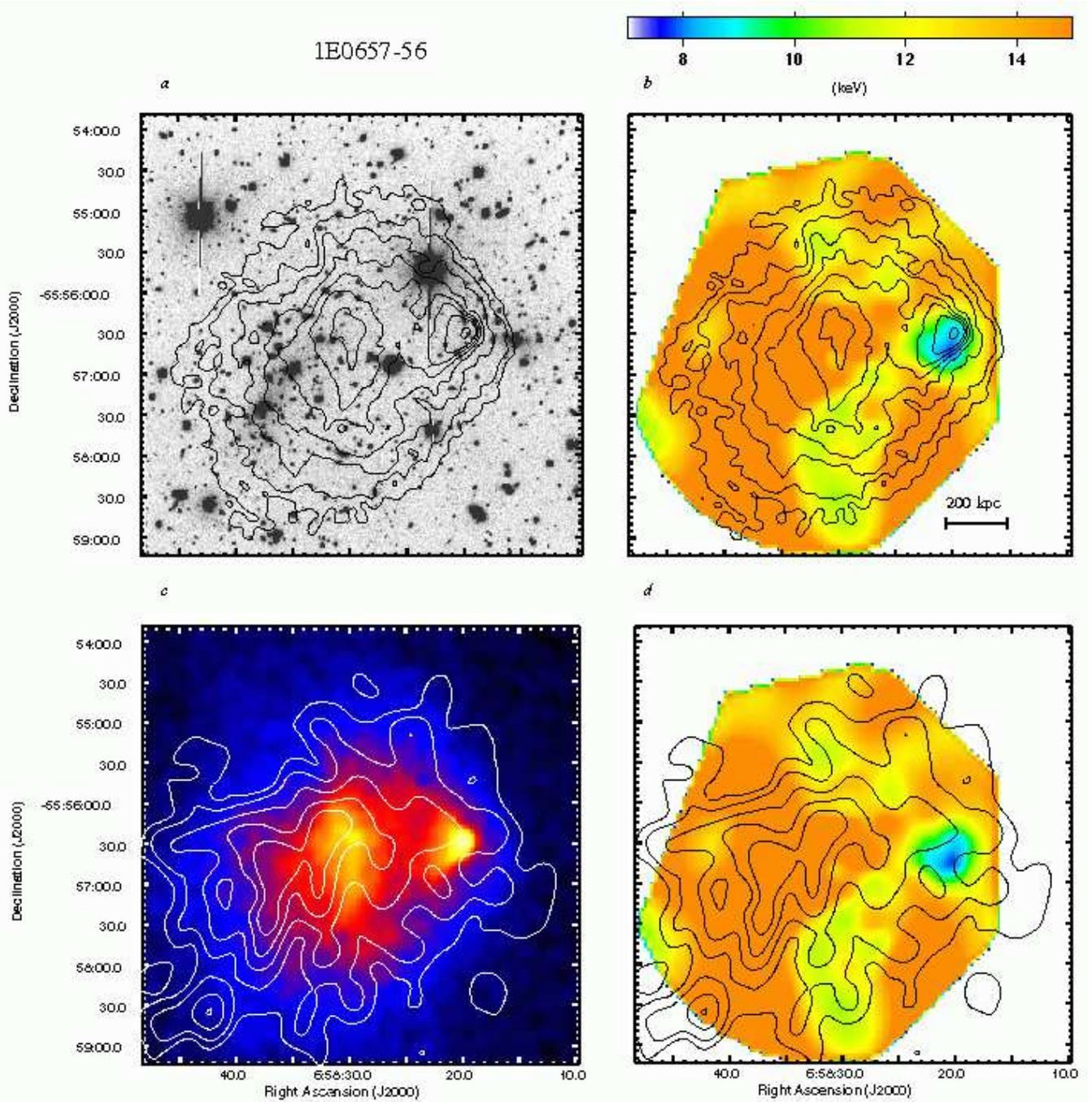}

\caption{1E0657-56. 
({\em a}\/):  X-ray contours overlaid on the optical R-band image
from the ESO New Technology Telescope (courtesy of E. Falco and
M. Ramella) image.  
The $0.8-4.0$ keV X-ray image is adaptively smoothed; contours are
 spaced by a factor of $\sqrt2$. 
({\em b}\/): X-ray contour plot overlaid on the temperature map (colors). 
({\em c}\/): The isocontour map at 1.3 GHz
  of the central region of 1E0657-56 (from Liang et al.\ 2000) overlaid on
  the X-ray image (colors). The 
  radio image has a FWHM of $24'' \times 22''$. The unrelated radio sources
  has been subtracted. The contour levels are: 
  (3 6 12 18 24)$\times$ $\sigma$, where the noise $\sigma$ is
   in the radio image is 90 $\mu$Jy/beam.
({\em d}\/): Radio contours overlaid on the temperature map (colors).
}
\label{fig:1E0657-56}
\end{figure*}

In Fig.\ \ref{fig:1E0657-56}, we compare the Liang et al. (2000) 1.3 GHz radio
image with a new, more accurate temperature map obtained from a deeper
\chandra\ exposure by Markevitch et al.\ (in preparation).  The new map and
the X-ray image suggest that the smaller subcluster has arrived at the
collision site from the southeast.  Most of its outer gas was shocked and
stripped during the collision with a bigger subcluster (whose largest
galaxies are now seen in the east). This stripped gas, together with the
shocked gas from the other subcluster, form the north-south bar-like
structure seen in the X-ray image, which is probably a pancake in
projection.  Shocks could not penetrate and stop the dense core of the
subcluster, and it is now continuing to the west, preceded by a bow shock
with $M\approx 3.5$. This subcluster should be generating vigorous
turbulence in its wake.

The radio halo peak is clearly offset from the X-ray brightness peak (in the
region that excludes the bullet), and instead is centered in the hottest
cluster region. The halo's eastern part is elongated along the presumed
infall trajectory of the subcluster.  Interestingly, the western part of the
halo extends all the way to the bow shock.

\section{Discussion and conclusions}
\label{sec:disc}

Substructure in the X-ray images and the galaxy spatial and velocity
distributions, as well as complex gas temperature structure, are signatures
of cluster mergers.  All these properties are common in clusters containing
radio halos. On the other hand, we do not know of extended radio halos in
relaxed clusters.  
Therefore, the available data suggest
that radio halos are related to mergers.

From the energetic grounds, mergers can indeed dissipate enough kinetic
energy --- simultaneously over a megaparsec-scale volume --- for the
maintenance of a radio halo.  However, it is not clear how exactly the
relativistic particles are accelerated.  In-situ acceleration (or
re-acceleration) of relativistic electrons during a merger can occur in
shocks (e.g. Sarazin 1999, 
Fujita \& Sarazin 2001) or in the gas turbulence (e.g.
Schlickeiser, Sievers, \& Thiemann 1987; Brunetti et al.\ 2001; 
Ohno, Takizawa, \& Shibata S. 2002, Fujita, Takizawa, \& Sarazin 2003).  

There are theoretical arguments against the shock hypothesis.  Most
importantly, a relatively strong shock with $M>4-5$ is believed to be
necessary for generation of an observable halo (e.g., Brunetti 2002).  Such
high Mach numbers should be very rare, as most merger shocks should have
$M\sim 1$ at the cluster center.
Gabici \& Blasi (2003) argued
that shocks expected in mergers of clusters with comparable masses are too
weak to result in significant non-thermal emission (the expected Mach number
increases with the mass ratio of the merging subclusters).  
Moreover, according to Miniati et al.\ (2001), such radio emission would
look more like radio relics at the cluster periphery rather than as radio
halos.  However, physics of collisionless shocks in clusters is not well
understood, and so at present even relatively weak merger shocks cannot be
completely ruled out as an acceleration mechanism.

Indeed, comparison of the radio halo and gas
temperature maps for a few merging clusters (MV01, M02a, M02b) 
hinted at their spatial
correlation, which could be easily explained if electrons were accelerated
in shocks.  A currently propagating or just-passed merger shock would leave
clear imprint in the cluster gas temperature maps in the form of hot
regions.  After the shock passage, regions of shock-heated gas quickly
expand adiabatically and come into pressure equilibrium with the
surroundings.  Large-scale gas motions during the merger subsequently mix
gases of different temperature, resulting in patchy temperature structure
(as seen in simulations, e.g., Roettiger, Stone, \& Burns (1999),
Takizawa 2000, Ricker \& Sarazin 2001) which persists for 
a considerable time
in the absence of thermal conduction (Markevitch et al.\ 2003a).  Any
relativistic electrons accelerated as the shock passes through the gas, will
be prevented from diffusing far from their origin by the magnetic fields
(the same fields that suppress thermal conduction and diffusion of thermal
electrons, e.g., Vikhlinin et al.\ 2001; Markevitch et al.\ 2003a),
and will follow the bulk motion of their host gas.  Thus, if relativistic
electrons are accelerated by merger shocks, and assuming a uniform magnetic
field (or, at least, a field uncorrelated with the gas temperature), one
expects strong spatial correlation between the radio halo brightness and the
hottest gas regions (in the absence of strong projection effects, of
course).

On the other hand, the gas turbulence, although not directly observable at
present, is expected to exist throughout the merging clusters, including
shock-heated and cooler gas regions (e.g., Sunyaev, Norman \& Bryan 2003).
Therefore in the turbulence scenario, there should be no strong correlation
between the radio brightness and the temperature.

The analysis presented here is a qualitative attempt to distinguish between
the shock and turbulence acceleration mechanisms by means of a systematic
comparison of radio halo maps with the temperature maps for a sample of halo
clusters with good radio and X-ray data.

All clusters studied here reveal clear signs of ongoing mergers and the
accompanying strong spatial temperature variations. Although in most of
these clusters, the merger geometry is ambiguous and the likely projection
effects complicate comparison with the radio data, we can draw several
conclusions.

We confirm the previously reported spatial coincidence of bright radio
features with the high temperature regions in A665, A2163, and \1e, which
was based on the shorter X-ray observations.  \1e\ and A520 are the best
examples of this temperature-radio connection.  This spatial ``coincidence''
is not quite a ``correlation' --- these clusters also exhibit bright radio
emission from some of the cool gas regions, but in the presense of
projection effects, exact correlation is not expected.  As discussed by
MV01, the leading edge of the ``limb-brightened'' (Jones \& Saunders 1996)
halo in A665 coincides with the apparent location of a possible relatively
strong shock ahead of the fast-moving cool core. This observation appear to
argue for the shock acceleration mechanism.

On the other hand, in A2319, which is at a very similar merger stage to
A665, we do not detect a particularly hot, shock-heated region in front of
the moving cool core, so its Mach number should be low.  Yet, the cluster
exhibits a radio halo whose brightest part follows rather closely the
distribution of the {\em cooler}\/ gas in the core (although at a lower
brightness level, the halo's NE-SW extension appears to coincide with the
outlying hotter regions). 

Moreover, in A754 and A773, the hottest cluster regions do not show radio
emission, at least at the present sensitivity, while there is radio emission
elsewhere in these clusters.  The same appears true for the small halo in
A2218, although the sensitivity of the radio data here is poor.  While radio
emission from cooler areas can be explained by projection effects, given
that the X-ray-derived projected temperature is biased toward the denser,
cooler gas on the line of sight, projection cannot explain the absence of
the radio emission from the hottest cluster regions.  We believe that this
is a strong observational argument against merger shocks, at least the
relatively weak ones expected in most mergers, as the main acceleration
mechanism.

This leaves turbulence as a more viable mechanism. Indeed, all clusters in
our sample where the merger geometry is tractable --- \1e, A520, A2319, and
A665 --- exhibit relatively small, dense, cool moving clouds which are
likely to generate turbulence in their wake.  The radio halos are observed
along the path of these moving clouds.  In A754, the eastern halo lies along
the bright cool X-ray tongue, which is likely to be the flow of dense gas
possibly generating turbulence at its interface. In A773, the passage of the
eastern mass peak through the gas of the main subcluster may also generate
turbulence in the bright radio area.  The remaining clusters have
sufficiently uncertain merger geometries to be consistent with the
possibility of strong turbulence in the right regions.

However, with turbulence alone, it is difficult to explain the observed
extension of the radio halo in \1e\ ahead of the gas bullet all the way to
the bow shock, a similar extension into the shock region ahead of the core
in A665, and possibly in A520.  Turbulence caused by these cores cannot
precede the fast-moving core, especially in such a high-velocity merger as
\1e. Thus, in systems like these, at least some of the relativistic
electrons should be accelerated in shocks.  Incidentally, \1e\ has a shock
with the highest-known Mach number ($M\approx 3.5$, M02a) and A665 appears
to contain a relatively high-$M$ shock as well ($M\approx 2$, MV01).

We conclude that in most clusters, the radio halo electrons are probably
accelerated by turbulence, but in those rare cases when shocks with $M\simeq
2-3$ are present, these shocks also appear to contribute in the
acceleration.  In this regard, we note that Kempner \& David (2003)
performed a similar radio--X-ray comparison for another halo cluster, A2744,
and concluded that both turbulence and shock acceleration may be present in
that cluster.

In conclusion, we note that maps of the spectral index of the radio halo
emission would be invaluable for pinpointing the electron acceleration sites
(because the halo spectrum steepens as the relativistic electrons lose
energy).  Such data would be especially illuminating in clusters such as
\1e, A665 (Feretti et al., in preparation), and A520, where both turbulence
and shocks are likely to be present and can be separated spatially.  Also,
for these clusters, higher-resolution halo images are required to determine
if the front edges of the halos indeed coincide with the shock fronts.

\acknowledgements

FG thanks the hospitality of the Harvard-Smithsonian Center for Astrophysics
where most of this work was done.  Support was provided by NASA contract
NAS8-39073, \chandra\ grants GO2-3164X and GO2-3165X, and the Smithsonian
Institution.  We thank M. Murgia for the use of the Synage++ program.

\vspace{4mm} 

NOTE ADDED IN PROOF: After this paper was submitted, an independently
derived \chandra\ temperature map for A2319 has been published by O'Hara,
Mohr, \& Guerrero (2004). It is in good agreement with our map.



\end{document}